# Differential Refractive index sensor based on Photonic molecules and defect cavities.

A. Andueza[1,a,c], J. Pérez-Conde[b] and J. Sevilla[a,c]

[a] Department of Electrical and Electronic Engineering, Universidad Pública de Navarra, Campus de Arrosadía, 31006, Pamplona, Spain.

[b] Department of Physics, Universidad Pública de Navarra, Campus de Arrosadía, 31006, Pamplona, Spain.

[c] Institute of Smart Cities, Universidad Pública de Navarra (UPNA) - 31006 - Pamplona, Navarra

**Abstract**

We present a novel differential refractive index sensor based on arrays of photonic molecules (PM) of dielectric cylinders and two structural defect cavities. The transmission spectrum of the photonic structure proposed as sensor shows a wide photonic stop band with two localized states. One of them, the reference state, is bound to a decagonal ring of cylinders and the other, the sensing state, to the defect cavities of the lattice. It is shown that defect mode is very sensitive to the presence of materials with dielectric permittivity different from that of the surrounding cylinders while the state in the PM is not affected by their presence. This behavior allows to design a device for sensing applications. A prototype of the sensor, in the microwave region, was built using a matrix of 3x2 PM arrays made of soda-lime glass cylinders ($\varepsilon_c$ = 4.5). The transmission spectra was measured in the microwave range (8-12 GHz) with samples of different refractive index inserted in the defect cavities. Simulations with Finite Integration time-domain Method are in good agreement with experiments. We find that the response of the sensor is linear. Device measurement range is determined by the dielectric constant of the cylinders that make up the device. The results here presented in the microwave region can be extrapolated to the visible range due to scale invariance of Maxwell equations. This make our prototype a promising structure as sensor also in the optical range.

*Keywords: Remote sensing and sensors, Photonic bandgap materials; Photonic crystals; Optical sensing and sensors.*

## 1. Introduction

Photonic crystals (PC) based on optical cavities offer many opportunities to create, control and design localized photonic states. The properties of these photonic states in the cavities are similar to those of confined electron states in atoms and we may refer to them as photonic atoms. Thus, it is possible to define more complex structures from "photonic atoms", termed photonic molecules (PM), formed by a cluster of several coupled photonic atoms [1].

It has been proved that a quasi-crystalline arrangement of dielectric scatterers can present well defined transmission frequencies inside a wide bandgap. The stop band is mainly due to global average order shown by the quasicrystal structure [2–4], while the transmission frequencies in the gap are generated short range structures of the quasicrystal and that resonate collectively

---

at these definite frequencies. Making an analogy of the cylinders as photonic atoms, these structures would be photonic molecules.

PM structures consist of two or more light-confining resonant cavities such as Fabry-Pérot resonators [5], microspheres [6], microrings [7], point-defect cavities in photonic crystals (PC) [8], microdisk [9,10], coupled-resonator optical waveguide (CROW) [11], etc. On the other hand, more complex photonic molecules can be created from coupled Photonic Crystal Cavities (PCC) [12]. These structures are dielectric defects in photonic crystals which generate strongly localized states in the Photonic Band Gaps (PBG) similarly to the case of impurities in atomic crystals [13]. PCC are a suitable medium for the control and confinement of light at scales of the order of a light wavelength in the material. The properties of PCC make them also strong candidates for efficient single photon generation for quantum information processing [14], development of light emitters [13], optical waveguides [13,15] or new nonlinear optics phenomena [16].

In addition, PM's can be used for sensing because they are very responsive to any modification of refractive index surrounding them, their shape and the refractive index of the resonant cavity. Typically, a sensor based on PM's is designed to detect the resonant shift in its frequency response produced by the change of the refractive index of the surrounding environment. Nowadays, there are many designs to sense and measure refractive index based on PM and defect cavities. For instance, photonic molecules have been proposed as sensing refractive index using dielectric microdisks [10]. Other sensor, such as microcavities in 2D arrays of PC [17], sandwich air holes [18], waveguides defect cavities [19], double hole defects [20], ring defects coupled resonators [21] and clusters of microparticles [22] are also successfully used to design and fabricate optical refractive index sensors.

In this paper we present a device that takes advantage of two of the above mentioned structures, PM and PCC in order to build a differential refractive index sensor. Our novel proposal of refractive index sensor combines the resonance effect produced in defect cavities and photonic molecules to design a differential sensor based on two modes, one of them sensitive to variation of refractive index: a state strongly dependent of refractive index of the sample and another totally independent of it. Employing these two modes we can design a differential sensor where the value of the refractive index would be determined by difference of the frequency value between each mode. Differential sensors provide significant advantages over simple sensors such as cross-sensitivities reduction, interference and noise cancelation, etc.

The proposed structure is a 2D photonic crystal made of dielectric cylinders. A building block containing a photonic molecule is repeated six times, in a 3x2 array. The structure holds two positions (where four of the building blocks meet) that act as defect cavities. The output of the sensor is the frequency value of two states in the gap. The frequency of one of the resonances varies linearly with the refractive index of the material inserted in the cavities. The frequency of the other state remains unchanged. A prototype of this sensor was built with glass cylinders and measured in the microwave region. Numerical calculations based on Finite Integrated Method (FIM) [23] were also performed for the same system and showed a very good agreement with the experimental measurements.

## 2. Experimental

### 2.1.- Sensor design

The structure is generated by the repetition of a building block made of 30 cylinders. The building block is taken from a periodic approximant used to generate a decagonal quasicrystal (QC) of dielectric cylinders [2]. This structure includes a decagonal ring in the center able to hold localized resonant states [2,3] therefore acting as a photonic molecule. One of these building blocks is indicated by a dashed (green) line in figure 1. Its structure is fully determined by the cylinder radius and the parameter *a*, because its height and width are given by:

$h=a(2+2\tau)$   and   $w=2a\left(\sqrt{1-\tau^2/4}+\sqrt{4-1/\tau^2}\right)$ , being $\tau$ the golden number $\tau=(1+\sqrt{5})/2$.

The building block is repeated six times forming a 3x2 array (of 191 cylinders) that forms two cavities in the spots where four of the building blocks meet. These cavities, labeled C1 and C2, are shown in figure 1 as solid (blue) circles. The simplest system based on this framework, a 2x2 array of building blocks (with four PMs and one cavity) was also analyzed (see supplementary material for details in Dataset 1, ref [24]), The results were considerably worse than those of the above-mentioned structure, so we focus in the 3x2 array.

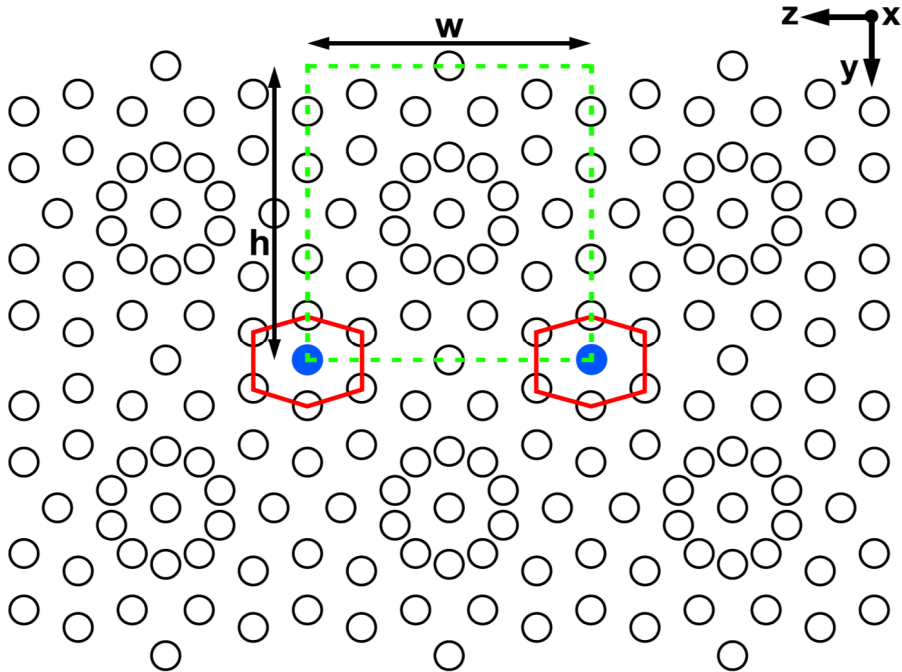

*Figure 1. Sensor design. Disposition of 191 cylinders structured as a 3x2 array of building blocks, depicted by a green dashed box. Blue filled circles correspond to the spots where samples of different refractive index will be inserted to be measured. These spots are in the defect cavities (red solid line) labeled C1 and C2.*

### 2.2. Prototype construction and characterization

A prototype of the sensor described in the previous section was fabricated with glass cylinders. The prototype is shown in figure 2. Soda-lime glass cylinders of 3 mm radius and 250 mm length were hold parallel to each other, fixed between two wooden pieces where their positions had been previously carved (with a numerical control drill). The decagonal rings acting as PM present an edge length *a*=12 mm with radius *r*=0.25*a*. The aspect-ratio (AR) which relates the length and the diameter of the cylinders, it is 42, approximately. The dispersion for radius of cylinders was also measured, found to be less than 0.2% of the nominal value, 3mm.

The cylinders dielectric permittivity at microwave frequencies was estimated at $\varepsilon_c$=4.5 as follows: first, we fabricated and measured the transmission spectra in perpendicular incidence, with TE (transversal electric) and TM (transversal magnetic) polarizations, of triangular and square regular lattices. Then we compared these results with simulations of the both structures and both polarization with the dielectric constant as only free parameter; the best fit was found for $\varepsilon_c$=4.5.

In order to analyze the performance of the prototype as a sensor, transmission spectra were measured after inserting cylinders of different materials (with different dielectric permittivity values) at the sensing positions (solid blue circles in figure 1).

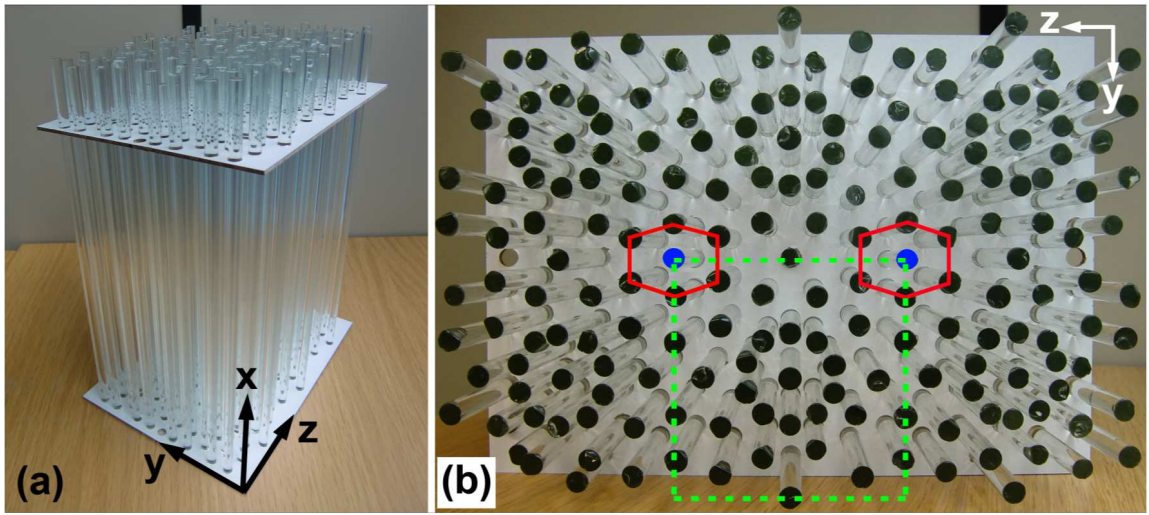

*Fig. 2. (a) Image of the built-up glass sensor completed with axis notation used in the text. (b) Detail of the fabricated sensor indicating the building block, defects cavities and cylinders samples*

Transmission spectra were measured using a vector network analyzer (VNA) HP 8722ES spliced to rectangular horn antennas 79mm × 46mm (Narda Model 640), aligned and separated approximately 400 mm from each other. The sample is placed in the space between the antennas, with the cylinders perpendicular to the incidence direction [25]. The measurement bandwidth is similar to antennas one and ranges from 8 GHz to 12 GHz. The cylinder and dielectric permittivity of the samples were selected in order to tune the modes in the bandwidth of the experimental setup. A differential measurement technique was used to register the transmission spectra. First, the transmission spectrum without the sample was recorded on the network analyzer. Next, the transmission spectrum of structure was measured. The analyzer automatically provided the difference between the spectra with and without the sample. A smoothing algorithm Savithzky-Golay from VNA (span for the moving average of 6 samples) was used to reduce the ripple noise of the measurements due to the finite size of the sensor.

In this work we consider the transverse magnetic (TM) polarization of the incident radiation, with the electric field vector parallel to the cylinders axis. The influence on the measurements of these drilled wooden pieces supporting the cylinders is negligible since the cylinders have a length of 25 cm and the antenna aperture never exceed 20 cm at 12 GHz.

Numerical calculations were carried out with CST MICROWAVE STUDIO™, a commercial code based on the Finite Integration time-domain Method (FIM) [23]. This program is an electromagnetic field sG:\LACIE\Memoria_PC\papers\Rosetas\manuscrito\obsoletoimulation software package especially suited for analysis and design in the high-frequency range. Considering the radiation traveling in the z direction, a Transverse Electric and Magnetic (TEM) mode with the electric field along *x*-axis and the magnetic field along *y*-axis was injected at normal incidence on the sensor. The symmetry of the sensor and the orientation of the electromagnetic field allow us to simplify the problem restricting the calculation to a unit cell consisting of one and a half consecutive building blocks. In our case two symmetry planes are considered, *xy* and *xz*. Every symmetry plane reduces the calculation time by a factor of 2.

### 3. Results and discussion

#### 3.1. Calibration curve

In the sensor prototype here proposed, the variable that encodes the measurand (refractive index of a sample) is the frequency of the sensing state. To ease the calibration of this sensing state, the sensor presents another state, the reference state, which remains unaltered for variations of the measurand. The calibration curve of the sensor was obtained by simulating the transmission spectrum of the structure for different values of the dielectric permittivity $\varepsilon_s$ of the samples inserted in the sensing spots. Two types of samples have been considered: (i) cylinders of the same radius of the rods composing the sensor ($r_2$=3 mm), and (ii) thinner cylinders ($r_2$=2.5 mm). The results are presented as 3D color maps in figure 3. The vertical axis represent the frequency (ranging from 9 to 12 GHz), the horizontal axis depicts the dielectric permittivity $\varepsilon_s$ of the samples (ranging from $\varepsilon_s$=1 to $\varepsilon_s$=5.5) while the transmitted intensity for each case is presented as the color of the points in the plot (and varies from 0 to -30 dB related to full transmission).

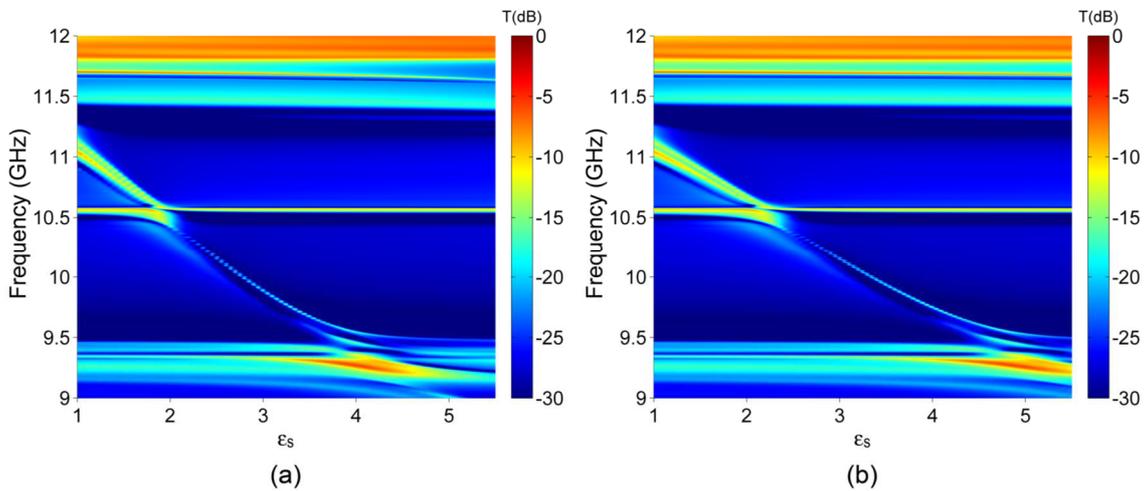

*Figure 3. Colormap plots of the calculated transmission spectra (dB) as a function of frequency (f) and dielectric permittivity ($\varepsilon_s$) of the samples at the cavities C1 and C2 with a radius of (a) $r_2$=3 mm and (b) $r_2$=2.5. Dielectric permittivity and length of the cylinders are $\varepsilon_c$=4.5 and l=25 cm, respectively. Horizontal yellow lines at 10.57 GHz correspond to reference mode of the sensor. Oblique yellow/cyan lines correspond to sensing mode. Vertical white-dashed line corresponds to experimental measurements.*

In Figure 3 we see that the sensing mode frequency decreases linearly with the mesurand value ($\varepsilon_s$). In contrast, the reference mode do not change for any dielectric permittivity value. The variation of the sensing state start at $\varepsilon_s$=1, the air value, corresponding to no sample in the sensing spots. Around $\varepsilon_s$=2 the sensing mode superimposes over the reference mode, generating region that could be difficult to interpret when measuring. This seems to be due to the mutual interaction between both modes when their frequency values coincide. The linear behavior of the sensing state ends when it approaches the band gap edge. This happens for $\varepsilon_s$ values close the sensor cylinders permittivity. However, the precise value of $\varepsilon_s$ depends on the measurand cylinders radius; this is the main difference between (a) and (b) in figure 3. For thicker measurand cylinders ($r$=3mm, figure 3a) the linear variation of the sensing state ends around $\varepsilon_s$=3.5, while in the case of thinner ones ($r$= 2.5 mm, fig 3b) the linear range extends up to $\varepsilon_s$=4.5. As the vertical span is the same for the two cases presented, extending the measuring range directly implies reducing the sensibility of the sensor. We can conclude that there is a compromise between measuring span and sensibility, and this compromise can be tuned with the radius of the measurand cylinders. The calculated value of refractive index sensitivity of the sensor, expressed in GHz per unit of refractive index (RIU), is 0.47 GHz/RIU for the first case ($r$=3mm, fig 3-a) and 0.4 GHz/RIU in the second ($r$= 2.5 mm, fig 3-b)

| Measurand radius r (mm) | Measurement range ($\varepsilon_s$) | Sensibility (GHz/RIU) |
|---|---|---|
| 3 | $1 - 3.5$ | 0.47 |
| 2.5 | $1 - 4.5$ | 0.40 |

Table 1. Summary of the main characteristics of the calibration curve in the two measuring conditions studied

The frequency change of sensing mode allows us to design a measurement process for the sensor where the main parameter is only the frequency shift of the sensing mode with respect of the reference mode. Besides, this feature can be very useful to calibrate the operation range of the sensor only adjusting the frequency where the reference mode is excited.

Changing the measurand cylinders radius the upper value of the measurement range can be varied, but au to a certain point. There is a limit imposed by the dielectric permittivity of the cylinders that compose the sensor body ($\varepsilon_c$=4.5 in our case). If we want to extend the operation range of the sensor we must increase the dielectric permittivity of the cylinders which form the sensor structure. This increase must reach the upper value of permittivity to be measured. In order to check this point, we calculated the transmission spectra for a sensor with cylinders of $\varepsilon_c$=7.5, $r/a$=0.25 and $r$=3 mm as a function of frequency and dielectric permittivity of the sample (see supplementary material for details in Dataset 1, ref. [24]). This study confirms our initial prediction: an appreciable increase in the measurement range can be achieved only by rising the dielectric permittivity of the cylinders of the structure.

### 3.2. Measured spectra

In order to validate simulated results, some measurements were made with the prototype of glass cylinders. Experimental transmission spectra were measured inserting into the center of the cavities C1 and C2 two pair of samples (see table 2). Also the transmission spectra without samples were recorded, this correspond to $\varepsilon_s$=1 for any radius. The cases with samples are marked in the calibration curves (figure 3) as vertical doted lines while the one without samples ($\varepsilon_s$=1) can be found in the left hand edge of the calibration curves.

| Experimental configuration | Sample | | Simulated results (Frequencies) | | | Measured results (Frequencies) | | |
|---|---|---|---|---|---|---|---|---|
| | Radius (mm) | $\varepsilon_s$ | Reference state (GHz) | Sensing state A (GHz) | Sensing state B (GHz) | Reference state (GHz) | Sensing state A (GHz) | Sensing state B (GHz) |
| 1 | 2.5 | 4.0 | 10.57 | 9.78 | Not visible | 10.70 | 9.69 | Not visible |
| 2 | 3 | 4.5 | 10.57 | Not visible | Not visible | 10.70 | Not visible | Not visible |
| 3 | Any | 1.0 | 10. 57 | 11.04 | 11.12 | 10.70 | 11.11 | 11.22 |

Table 2. Summary of the performed measurements

The results, together with the simulations of the same structures, are presented in figures 4 (measurements 1 and 3) and figure 5 (measurements 2 and 3). In general the correspondence between measurements and simulations is quite good. The spectra show the transmission stop band from 9.5 GHz to about 12 GHz, indicated with vertical dashed lines in the figures. Inside the stop band we can observe sharp and distinct experimental peaks. Although the peaks are sharper in the calculated results, the correspondence with the observed ones is remarkable.

One of the states is present in all the measurements, this is the reference state and has been labeled R. It is interesting to note that the other state, the one that is dependent on the $\varepsilon_s$ value, is really composed of two different peaks almost overlapping. This is particularly evident in the case without sample (labeled measurement 3), where the two peaks can be seen both, in the measurement and the calculation (labelled in the figures $S_{3A}$ and $S_{3B}$).

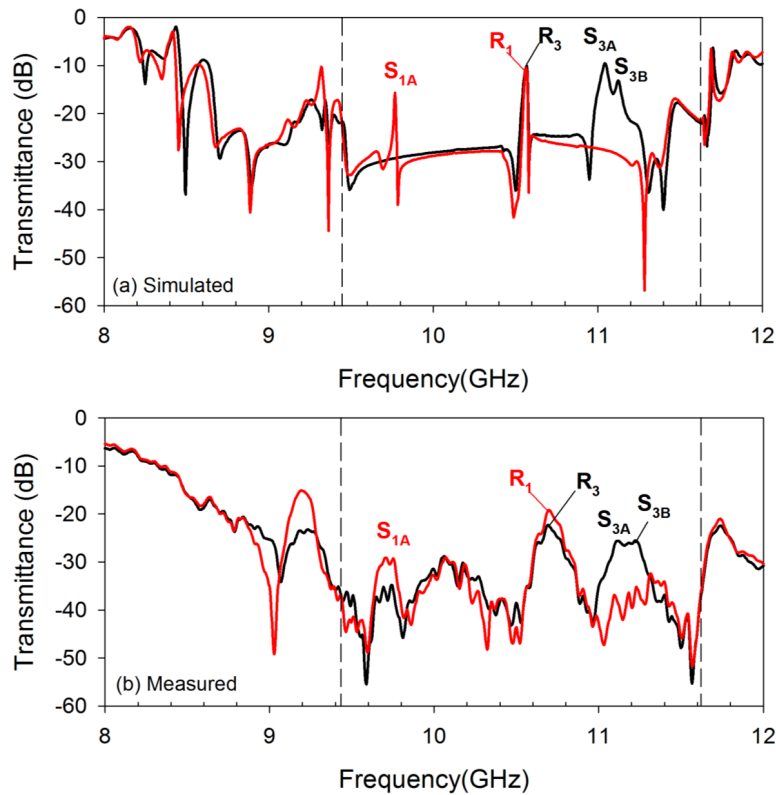

*Figure 4. (a) Calculated and (b) measured transmission spectra for the experimental configurations 3 (black solid line) and 1 (red dashed line). Resonances corresponding to the sensing (S marked) and reference modes (R marked) are labeled for each case.*

In the case of sample 2 (figure 5) the sensing state is not detected. This happens because the shifting in the state frequency value is big enough to move it outside the gap, making it indistinguishable of the gap edges as can be observed in the calibration curve (figure 3a).

The frequency values of the measured and simulated peaks are summarized in table 2. The difference between the measured and the calculated values is around 0.1 GHz (0.13 and 0.09 in the reference and sensing states respectively). The relative error in frequency is around 1 %., The results can be considered an excellent proof of concept of the working principle of the proposed sensor taking into account the fabrication tolerances of the prototype.

The influence of AR of the cylinders was analyzed (see supplementary material for details in Dataset 1, ref. [24]). The aspect-ratio does not produce significant variations in the performance of the sensor when it is larger than 20. Therefore, this result validates the choice of cylinders with length of 25 cm (AR of 42).

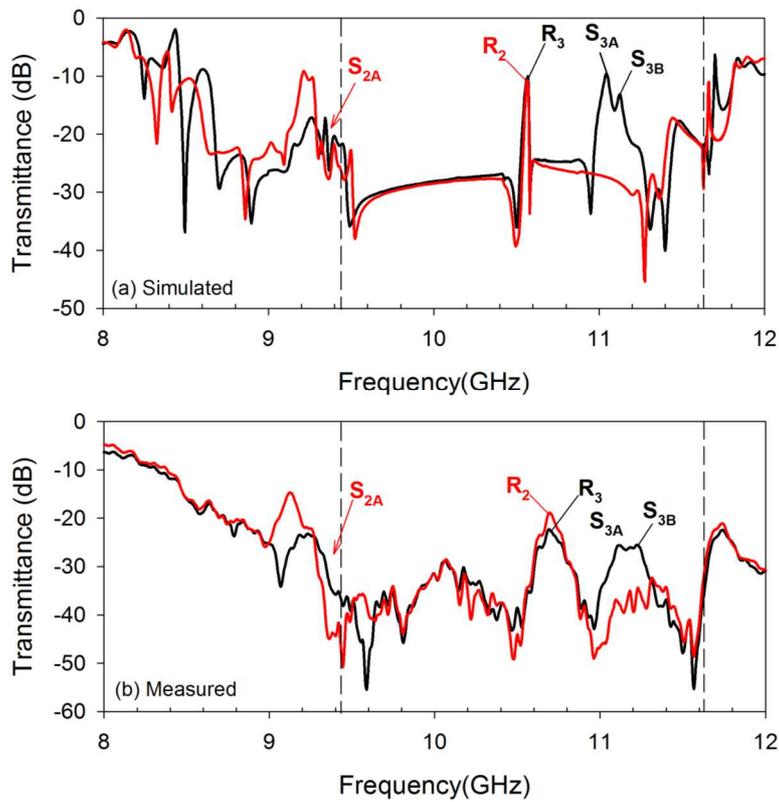

*Figure 5. (a) Calculated and (b) measured transmission spectra for the experimental configurations 3 (black solid line) and 2 (red dashed line). Resonances corresponding to the sensing (S marked) and reference modes (R marked) are labeled for each case. Vertical dashed lines correspond to the edge of the stop band.*

### 3.3. Field distribution of the resonant states

The results suggest that two resonance phenomena occurs simultaneously, exciting the reference and sensing modes. Also, it has been found that the sensing mode is twofold in some cases at least. In order to assess the origin of these modes we explore the calculated electric field distributions of the system. We compute the spatial distribution of the electric field of the modes in the gap seen in figure 4. For the calculation, polarized radiation in the x direction with unit amplitude propagates in the z direction. The obtained electric field distributions are depicted in figure. 6.

Electric field distributions of modes $R_1$ [figure 6(a)] and $R_2$ [figure 6(b)] are very similar, with the electric field localized mainly inside the cylinders; with the exception of the cylinders along the z axis where the electric field is located between them. This field pattern was previously reported [2,3] and corresponds to a localized state presents in a decagonal quasicrystal. Therefore we can conclude that the reference modes $R_1$ and $R_2$ are excited by the photonic molecules from the decagonal ring. The electric field distribution of sensing modes $S_{3A}$ and $S_{3B}$ (at 11.04 GHz and 11.12 GHz) are plotted in figure 6(b) and 6(c), respectively, while figure 6(e) corresponds to the sensing mode $S_1$ at 9.9 GHz. It is important to note that the electric field pattern in $S_3$ is strongly localized and enhanced on each cavity and presents even or odd symmetry along y-axis for the A and B states respectively. These kind of bonding and antibonding states in photonic crystals has been previously reported after the introduction of an air defect into the crystal [26,27]. In the case of $S_1$ (found at 9.9 GHz and corresponding to the case of a sample with $\varepsilon_c$=4) the electric field distribution is similar to the $S_3$ case, but only the bonding state is found. The antibonding mode is not excited when the refractive index of the sample cylinders has increased.

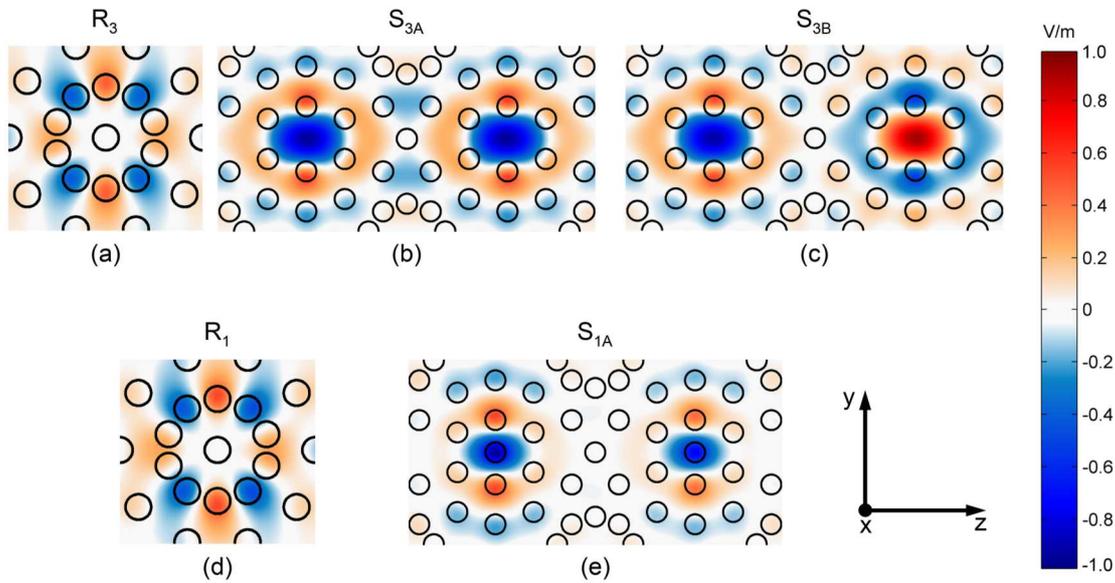

*Figure 6. Distribution of the electric-field for yz plane (a) of $R_1$ at f =10.57 GHz, (b) of S1 at f = 11.04 GHz and (c) $S_2$ at f =11.12 GHz without a sample. (d) Distribution of the electric-field for yz plane of $R_2$ at f = 10.57 GHz and (e) $S_3$ at f =9.78 GHz for a sample of $r_2$ = 2.5 mm and $\varepsilon_c$ =4. The red and blue colors indicate the electric field intensity and polarity.*

These results prove that this photonic structure presents sufficient capability to operate as refractive index sensor. In addition, the existence of a reference state allows the design of a differential version of the sensor. This differential operation provides multiple advantages such as systematic error cancelation, cross sensitivities reduction, etc. We have also shown that system parameters, such as size and dielectric permittivity of cylinders sets, structures and samples, can be fitted in order to optimize its response for a given measurement range. As the Maxwell equations governing the studied system are scale invariant, the results in the microwave regime can readily be extrapolated to the optic range. This transformation implies that the size of the materials and the wavelength are scaled with the same factor. Accordingly, the dielectric permittivity in the microwave range must be considered in the optical range, taking into account the imaginary part (which represents material losses) that can be neglected in the microwave range but rarely in optics. Materials as Titania ($TiO_2$), Alumina ($Al_2O_3$) or Silicon Carbide (SiC) could be used for THz and FIR, or Silicon (Si) and Zinc Sulfide (ZnS) for NIR and visible zones.

## 4. Conclusions

We have designed and fabricated a differential refractive index sensor prototype based on 3x2 photonic molecules array of dielectric cylinders containing two defect cavities. The building block is based on a quasi-crystaline approximant including a decagonal ring of cylinders acting as photonic molecule. This structure creates a stop band where two modes can be excited: the first mode is due to the PM (reference mode) and the second mode is due to the defect cavities (sensing mode). When the defect cavity is disturbed by a sample cylinder with different dielectric constant to the surrounding, the sensing mode suffers a frequency shift proportional to its permittivity value. However, the reference mode associated to the photonic molecule does not undergoes a frequency change and remains unaltered. Therefore, the dielectric permittivity can be calculated as the frequency difference between both resonances.

Transmission spectra for two samples with different dielectric permittivity were experimentally obtained in microwave region for a sensor fabricated with glass cylinders of permittivity $\varepsilon s = 4.5$ and $r = 3$ mm. They were in good agreement with the results of the numerical calculation of the sensor. The sensor presents a linear dependence with the refractive index of the sample in the cavity as well as a sensitivity of 0.47 GHz/RIU and 0.4 GHz/RIU for two samples with $r = 3$ mm and $r = 2.5$ mm, respectively. We found that the uncertainty of the sensor is lower than 5% for the experimental measurement. However, the span of the sensor was limited by the dielectric constant of the PM cylinders. The range of measurement can be expanded increasing the dielectric permittivity of the surrounding cylinders.

We believe that this device is well suited to design and develop a variety of novel refractive index optical sensors, as the scale invariance of electromagnetic waves allows the extrapolation of this sensor to the optical and nanoscale region as well.

## Acknowledgements

This work has been supported from Ministerio de Economía y Competitividad of Spain via Project TEC2014-51902-C2-2-R. The authors would also like to thank Santiago de Miguel for his technical support.